\documentclass[12pt]{article}
\usepackage{a4wide}
\usepackage{amssymb}
\usepackage{graphicx}
\usepackage{xcolor}
\begin{document}
{\renewcommand{\thefootnote}{\fnsymbol{footnote}}
\begin{center}
{\LARGE  Lessons for loop quantum gravity\\[2mm] from emergent modified gravity }\\
\vspace{1.5em}
Idrus Husin Belfaqih,$^1$\footnote{e-mail address: {\tt i.h.belfaqih@sms.ed.ac.uk}}
Martin Bojowald,$^2$\footnote{e-mail address: {\tt bojowald@psu.edu}}\\
Suddhasattwa Brahma,$^1$\footnote{e-mail address: {\tt sbrahma@exseed.ed.ac.uk}}
and Erick I.\ Duque$^2$\footnote{e-mail address: {\tt eqd5272@psu.edu}}
\\
\vspace{0.5em}
$^1$Higgs Centre for Theoretical Physics, School of Physics \& Astronomy,\\
University of Edinburgh, Edinburgh EH9 3FD, Scotland, UK\\
\vspace{0.5em}
$^2$ Institute for Gravitation and the Cosmos,\\
The Pennsylvania State
University,\\
104 Davey Lab, University Park, PA 16802, USA\\
\vspace{1.5em}
\end{center}
}

\setcounter{footnote}{0}

\begin{abstract}
  Most of the potential physical effects of loop quantum gravity have been
  derived in effective models that modify the constraints of canonical general
  relativity in specific forms. Emergent modified gravity evaluates important
  conditions that ensure the existence of a compatible geometrical space-time
  interpretation of canonical solutions, as well as phase-space covariance
  under different choices of canonical variables. This setting, specialized to
  modifications suggested by loop quantum gravity, is therefore an important
  contribution to physical evaluations of this approach to quantum
  gravity. Here, it is shown that emergent modified gravity restricts several
  ambiguities that existed in previous formulations, rules out several
  specific candidates, and provides a unified treatment of different types of
  (holonomy) modifications that had been thought to be physically
  distinct.
\end{abstract}

\section{Introduction}

Models of black holes can be constructed with relative ease in spherically
symmetric reductions of general, modified, or quantum gravity. Loop quantum
gravity is a popular area for such endeavors, which has given rise to a
growing set of inequivalent descriptions that are argued to descend from one
and the same quantum theory of gravity. Some of the differences between
competing proposals can be attributed to the fact that the originating theory
of full quantum gravity remains incompletely understood and subject to
quantization ambiguities. At present, it is impossible to construct black-hole
models directly within loop quantum gravity. Instead, some of its
characteristic ingredients, such as the use of different kinds of spatial
discreteness or the use of holonomies instead of connection or curvature
components, are modeled in more accessible equations of reduced models. The
choice of these ingredients and their implementation is not based on
derivations directly from the full theory and implies a certain measure of
ambiguity.

A question of greater concern is about the relationship between the resulting
equations and solutions of some abstract theory, and space-time properties
such as curvature or horizons that are essential for a definition and
interpretation of black holes. In loop quantum gravity, the underlying quantum
formulation does not reveal a space-time geometry in which relevant
calculations could be made. The theory, in its most reliable formulation, is
canonical, and therefore presents constraint operators that quantize the
classical generators of hypersurface deformations in space-time. However, the
quantization procedure modifies the classical generators by including, in a
semiclassical or effective description, fluctuation terms as well as
higher-order dependencies on some of the canonical variables as a trace of
spatial discreteness in a series expansion. Unless such modifications are
chosen carefully, modified generators have no relationship whatsoever with
hypersurface deformations or transformations of space-time. And even in
careful choices, the compatible space-time geometry, if it exists, is not
guaranteed to be immediately given by the gravitational phase-space variables
as in the classical theory, where some of the basic variables are usually
given by the spatial metric or a triad.

In most black-hole models of loop quantum gravity constructed so far, there is
no attempt to derive a space-time metric compatible with modified
constraints. Instead, the classical identification of metric components among
phase-space variables is used, even while the underlying
hypersurface-deformation generators are modified. However, as a general
property of (pseudo-)Riemannian space-time geometry, commutators of
hypersurface deformations depend on the metric
\cite{DiracHamGR,Katz,ADM,HypDef}, and therefore these two ingredients are not
independent of each other. This lacuna implies another, more worrisome source
of ambiguities because different gauge choices that, without modifications,
would imply different slicings of one and the same classical space-time
geometry, may easily lead to inequivalent predictions if quantum terms are
included. In many cases, no attempt has been made to change the gauge and
check whether there is any invariance. Such models of space-time and black
holes are unreliable. They also suffer from missed opportunities because
checking consistency conditions on the proposed space-time geometries suggests
powerful ways to identify viable models and narrow down the initial large set
of ambiguities.

We perform this task in the present paper, using the framework of emergent
modified gravity \cite{Higher,HigherCov,SphSymmMinCoup} that, compared with
models of loop quantum gravity, is based on closely related ingredients such
as a canonical formulation and related phase-space variables without an
extension of the classical phase space. Its main new ingredient is a complete
characterization of covariance conditions within the canonical theory,
combined with an unambiguous derivation of a compatible space-time metric for
a given set of consistent modifications of the gravitational constraints. We
will show that none of the black-hole models of loop quantum gravity discussed
in detail in recent work have a compatible space-time geometry. However, as
first shown by the specific cases in \cite{SphSymmEff,SphSymmEff2}, it is
possible to add symmetry-restoring terms that maintain some of the
modifications proposed earlier and make them compatible with a space-time
geometry in which curvature and horizons can be defined unambiguously. Other
forms of modifications can be ruled out as incompatible with space-time
covariance, reducing the amount of ambiguities in black-hole models.

Moreover, different versions of the compatible models, previously thought to
be physically inequivalent, can be shown to be related to each other by
canonical transformations, provided all of them are amended by
symmetry-restoring terms in order to make them compatible with space-time
covariance \cite{HigherCov}. Solutions derived in different gauges of a given modified theory
are then related by coordinate transformations of a compatible space-time
metric, a highly non-trivial condition that has not been met in many older
constructions.  As a result, we arrive at a new understanding of covariant
holonomy modifications in models of loop quantum gravity that combines
ingredients from the traditional classes of $\mu_0$ and $\bar{\mu}$-type
schemes for the phase-space dependence of holonomy terms. In particular, we
will demonstrate that $\mu_0$ and $\bar{\mu}$-type holonomy terms in a
covariant model can always be transformed into each other by a canonical
transformation. These two traditional schemes or possible alternatives with
different scale dependences therefore do not necessarily represent distinct
physical effects. Instead, there are additional modification functions in a
covariant Hamiltonian constraint that must be taken into account in order to
parameterize different outcomes analogous to the traditional distinction
between $\mu_0$ and $\bar{\mu}$.

We summarize relevant properties of spherically symmetric models of loop
quantum gravity and emergent modified gravity in
Section~\ref{s:Summary}. Section~\ref{s:Implications} then presents detailed
discussions of various implications of the latter framework for the former
setting, most of which have not been considered before. In particular, we
demonstrate the existence of a unified and covariant formulation of holonomy
modifications in Section~\ref{s:Schemes}, where we show canonical
relationships between the traditional $\mu_0$ and $\bar{\mu}$-schemes. We
conclude in Section~\ref{s:Conclusions}, where we also apply our results in
brief discussions of recent black-hole and collapse models in loop quantum gravity.

\section{Summary of spherically symmetric models of loop quantum gravity}
\label{s:Summary}

Spherically symmetric space-time solutions have a line element of the form
\begin{equation} \label{ds}
  {\rm d}s^2= -N^2{\rm d}t^2+q_{xx}({\rm d}x+M{\rm d}t)^2+
  q_{\vartheta\vartheta}({\rm d}\vartheta^2+\sin^2(\vartheta){\rm d}\varphi^2)
\end{equation}
with the lapse function $N(t,x)$, the radial shift vector component $M(t,x)$,
the radial metric $q_{xx}(t,x)$ and the angular metric component
$q_{\vartheta,\vartheta}(t,x)$. In this section, we review features relevant
for models of loop quantum gravity, as originally provided in
\cite{SymmRed,SphSymm,SphSymmHam}. 

\subsection{Triad variables}

For models of loop quantum gravity, it is
convenient to transform the spatial metric components to the components $E^x$ and
$E^{\varphi}$ of a spatial densitized triad,
\begin{equation} \label{E}
  E^a_i\tau^i \frac{\partial}{\partial x^a}=E^x\tau_3\sin(\vartheta)\frac{\partial}{\partial
    x}+
  E^{\varphi}\left(\tau_1\sin(\vartheta)\frac{\partial}{\partial\vartheta}+\tau_2\frac{\partial}{\partial\varphi}\right) 
\end{equation}
with ${\rm su}(2)$-generators $\tau_i$ used to separate the triad
components. Evaluating
\begin{equation}
  \delta^{ij}E^a_iE^b_j=q^{ab}\det(q_{cd})
\end{equation}
and
\begin{equation}
  \det(q_{cd})=|\det(E^a_i)|
\end{equation}
in terms of the components implies the relationships
\begin{equation}
  q_{xx}=\frac{(E^{\varphi})^2}{|E^x|} \quad\mbox{and}\quad
  q_{\vartheta\vartheta}=|E^x|\,.
\end{equation}
Since
\begin{equation}
  \det(E^a_i)= E^x(E^{\varphi})^2\sin^2\vartheta\,,
\end{equation}
the sign of $E^x$ determines spatial orientation through the handedness of the
densitized triad. We will assume right-handed triads throughout, using
$E^x>0$. (The sign of $E^{\varphi}$ does not have physical relevance as it can
be changed by a remnant ${\rm U}(1)$-gauge transformation given by conjugation
of (\ref{E}) by $\exp(2\pi\tau_3)=-\tau_3$.)

The momenta canonically conjugate to the metric or triad components are
classically related to extrinsic curvature of a spatial slice within
space-time. As a 1-form, a generic spherically symmetric extrinsic-curvature
tensor has the form
\begin{equation}
  K_a^i\tau_i{\rm d}x^a= K_x\tau^3{\rm d}x+ K_{\varphi}\tau^1{\rm d}\vartheta+
  K_{\varphi}\tau^2\sin(\vartheta){\rm d}\varphi
\end{equation}
with two independent components, $K_x$ and $K_{\varphi}$. The components
$E^a_i$ and $K_a^i$ are canonically related according to the Liouville 1-form
\begin{equation}
  \frac{1}{8\pi G} \int{\rm d}^3x K_a^i \delta E^a_i
\end{equation}
in Schwinger-type variables \cite{Schwinger},
with the exterior derivative $\delta$ on field space. Inserting spherically
symmetric components and integrating over the angles, this 1-form is turned
into
\begin{equation}
  \frac{1}{2G}  \int{\rm d}x
  \left(K_x\delta E^x+2K_{\varphi}\delta E^{\varphi}\right)
\end{equation}
and implies the Poisson brackets
\begin{equation}
  \{K_x(x),E^x(y)\}=2G\delta(x-y) \quad\mbox{and}\quad
  \{K_{\varphi}(x),E^{\varphi}(y)\}=G\delta(x-y)\,.
\end{equation}
The remaining factor of two can be removed by working with the momenta
$P_x=\frac{1}{2}K_x$ of $E^x$ and $P_{\varphi}=K_{\varphi}$ of
$E^{\varphi}$.

In these variables, the diffeomorphism and Hamiltonian
constraints are given by
\begin{equation}
  D[M]= \frac{1}{G}\int {\rm d} x M \left( P_\varphi' E^\varphi - P_x
    (E^x)' \right)
\end{equation}
and
\begin{eqnarray} \label{Hclass}
    H [N] &=& \frac{1}{G} \int {\rm d} x\ N \left(- \frac{E^\varphi}{2 \sqrt{E^x}} P_\varphi^2
      - 2  \sqrt{E^x}P_xP_\varphi \right.\\
    &&\qquad\qquad\quad\left.- \frac{E^\varphi}{2 \sqrt{E^x}}+
              \frac{((E^x)')^2}{8 \sqrt{E^x} E^\varphi}
     - \frac{\sqrt{E^x} (E^x)' (E^\varphi)'}{2 (E^\varphi)^2} 
    + \frac{\sqrt{E^x} (E^x)''}{2 E^\varphi}
    \right)\,. \nonumber
\end{eqnarray}

\subsection{Types of modifications}
\label{s:Types}

Loop quantum gravity constructs operators for the canonical geometrical
variables and the constraints by using holonomies of a spatial
${\rm SU}(2)$-connection instead of extrinsic curvature \cite{LoopRep}. This
procedure modifies in particular the Hamiltonian constraint because terms
quadratic in the momenta cannot be directly represented by
${\rm SU}(2)$-holonomies, which have bounded components. The quantum
constraint, modeled on proposals such as \cite{AnoFree} in the full theory or
\cite{SphSymmOp} in spherically symmetric models, is therefore a quantization
not of (\ref{Hclass}) but rather of an expression in which the momentum-dependent
terms have been modified to certain non-quadratic functions of $P_x$ and
$P_{\varphi}$ together with the original $E^x$ and $E^{\varphi}$. Since there
are two momentum-dependent terms in (\ref{Hclass}) and two independent momenta,
several options are available for modifications.

The most general anomaly-free version with a derivative structure as in the
classical theory has been derived in \cite{JR} and discussed further in
\cite{ConsCosmo} (as well as \cite{SphSymmMaxwell} in a special case), given
by a Hamiltonian constraint of the form
\begin{eqnarray} \label{Hmod}
    H [N] &=& \frac{1}{G} \int {\rm d} x\ N \left(-\alpha(E^x) \frac{E^\varphi}{2
              \sqrt{E^x}} f_1(P_{\varphi},P_x,E^x,E^{\varphi}) 
      - 2 \bar{\alpha}(E^x) \sqrt{E^x}f_2(P_\varphi,P_x,E^x) \right.\nonumber\\
    &&\qquad\qquad\quad- \alpha_{\Gamma}(E^x)\left(\frac{E^\varphi}{2 \sqrt{E^x}}-
              \frac{((E^x)')^2}{8 \sqrt{E^x} E^\varphi}\right)\\
&&  \qquad\qquad\quad\left.   - \bar{\alpha}_{\Gamma}(E^x)\left(\frac{\sqrt{E^x} (E^x)' (E^\varphi)'}{2 (E^\varphi)^2} 
    - \frac{\sqrt{E^x} (E^x)''}{2 E^\varphi}\right)
    \right) \nonumber
\end{eqnarray}
with the conditions that
\begin{equation}
  f_2(P_{\varphi},P_x,E^x)=P_x F_2(P_{\varphi},E^x)
\end{equation}
and
\begin{equation} \label{alphaf}
  \left(\bar{\alpha}\alpha_{\Gamma}-
    2E^x\left(\frac{\partial\bar{\alpha}}{\partial E^x}\bar{\alpha}_{\Gamma}-
      \bar{\alpha} \frac{\partial\bar{\alpha}_{\Gamma}}{\partial
        E^x}\right)\right)F_2+ 2\bar{\alpha}\bar{\alpha}_{\Gamma} E^x
  \frac{\partial F_2}{\partial E^x}= \frac{1}{2} \alpha\bar{\alpha}_{\Gamma}
  \frac{\partial f_1}{\partial P_{\varphi}}\,.
\end{equation}
If these conditions are fulfilled, the structure function in the Poisson
bracket of two Hamiltonian constraints is multiplied by
\begin{equation} \label{beta}
  \beta=\bar{\alpha}\bar{\alpha}_{\Gamma} \frac{\partial F_2}{\partial
    P_{\varphi}}\,.
\end{equation}
(A version of these equations with $\alpha=1=\bar{\alpha}$ was recently
rederived as a Lemma in \cite{LTBPol}.)  It then follows that $f_1$ cannot
depend on $P_x$ or $E^{\varphi}$. An extended version with additional
derivative terms that do not appear in the classical Hamiltonian constraint
but are of the same order have been derived more recently in
\cite{HigherCov,SphSymmMatter,SphSymmMatter2}. We will use (\ref{Hmod}) for
specific examples in this section and return to the complete expression in
Section~\ref{s:EMG}. We will also discuss how anomaly-freedom, the general
condition imposed in (\ref{Hmod}) is necessary but not sufficient for general
covariance, even though it is restrictive and already rules out some models
considered in loop quantum gravity.

In models of loop quantum gravity, the specific form of
${\rm SU}(2)$-holonomies suggests that the modification functions should have
specific periodicity properties. Such features could be directly implemented
by assuming periodic functions $f_1$ and $F_2$. Sometimes, it is also
argued that periodicity should rather be obtained in a combination such as
$P_{\varphi}/\sqrt{E^x}$ that takes into account a possible dependence of the
holonomy length on the canonical metric variables in the sense of lattice
refinement \cite{InhomLattice,SchwarzN}. Again, several versions exist for these
dependencies within the limits imposed by conditions such as (\ref{alphaf}).

In particular, a dependence of $f_1$ and $F_2$ on $P_{\varphi}$ and $E^x$ is
such that strictly periodic behavior, for instance
$F_2(P_{\varphi})=\delta^{-2} \sin^2(\delta P_{\varphi})$, is possible only
with constant $\delta$. If this constant is replaced by a function of $E^x$,
as in $F_2(P_{\varphi},E^x)=\delta_2^{-2}E^x\sin^2(\delta_2 P_{\varphi}/\sqrt{E^x})$,
the partial derivative by $E^x$ in (\ref{alphaf}) produces a non-periodic term
proportional to $P_{\varphi}$, which must then contribute to $f_1$. An example
of this form can be found in \cite{LTBPolII}, where
\begin{equation}
  F_2(P_{\varphi},E^x)=\sqrt{E^x} \frac{\sin(2\delta_3
    P_{\varphi}/\sqrt{E^x})}{2\delta_3} 
\end{equation}
and
\begin{equation}
  f_1 = \frac{3E^x}{\delta_3^2} \sin^2(\delta_3P_{\varphi}/\sqrt{E^x})-
  \frac{\sqrt{E^x}P_{\varphi}}{\delta_3} \sin(2\delta_3P_{\varphi}/\sqrt{E^x})\,.
\end{equation}
For later reference, we note the momentum dependence
\begin{equation} \label{momentumterms}
  \frac{3}{\delta_3^2} \sqrt{E^x}E^{\varphi}
  \sin^2(\delta_3P_{\varphi}/\sqrt{E^x})+ \frac{2E^xP_x-
  E^{\varphi}P_{\varphi}}{\delta_3} \sin(2\delta_3P_{\varphi}/\sqrt{E^x})
\end{equation}
of the Hamiltonian constraint in this case.

A version of a modified constraint that does not obey (\ref{alphaf}) has been
given in \cite{MimeticCGHS}: 
\begin{equation}
  \frac{1}{2\delta_4^2}\sqrt{E^x}E^{\varphi}\left(\sin^2(2\delta_4\sqrt{E^x}P_x/E^{\varphi})-
    4\sin^2(\delta_4\sqrt{E^x}P_x/E^{\varphi}+
    {\textstyle\frac{1}{2}}\delta_4P_{\varphi}/\sqrt{E^x})\right)\,.
\end{equation}
Here, using a Taylor expansion in $P_x$ (which is sufficient because anomalies
must cancel out order by order in $P_x$), a $P_x^2$-term cancels out only for
$\delta_4=0$ but not for non-zero values with modified constraints. Such a
Hamiltonian constraint is not anomaly-free because they violate the condition
that any anomaly-free (\ref{Hmod}) must be linear in $P_x$.

\subsection{Consistency conditions}
\label{s:Consistency}

The introduction of periodic functions serves two main purposes. First, it
facilitates quantization by using bounded holonomy operators for terms in the
modified Hamiltonian constraint. Secondly, it can be viewed as a specific way
of defining new theories of modified gravity with physically interesting
space-time solutions that may have bounded curvature. In the latter case, the
dynamics is determined by Hamilton's equations generated by the modified
constraints, implying constrained dynamics subject to strong consistency
conditions that are unlikely to be met by unrestrained modifications. Most
modifications of this type that have been proposed in the literature do not
consider such consistency conditions but rather assume a fixed gauge or
introduce matter fields in a preferred reference frame by
deparameterization. It is then assumed, often implicitly, that a space-time
line element of the classical form (\ref{ds}) can still be used to equip
canonical solutions for $E^x$ and $E^{\varphi}$ with geometrical meaning.

The fallacy of this assumption can easily be seen by an application of a
simple canonical transformation $\tilde{E}^{\varphi}= 2E^{\varphi}$ and
$\tilde{P}_{\varphi}= \frac{1}{2}P_{\varphi}$ while keeping $E^x$ and $P_x$
unchanged. Any solution for $E^{\varphi}$ after the transformation can be
obtained by multiplying the old solution by two. If, according to a standard
assumption, the new $\tilde{E}^{\varphi}$ (of the ``modified'' theory) is used
to define a spatial metric $(\tilde{E}^{\varphi})^2/E^x=4q_{xx}$ to be used in
the space-time line element (\ref{ds}) instead of the original $q_{xx}$, this
metric component is multiplied by four. A constant factor of $q_{xx}$ can be
absorbed in the $x$-coordinate, but this will change any $x$-dependence of
$E^x$. According to this procedure, therefore, canonical transformations do
not preserve space-time geometries constructed from the solutions.

This argument shows that a compatible space-time metric must be derived from a
canonical theory in order to provide a reliable geometrical interpretation of
the solutions, which, like any physical predictions, should not depend on canonical
transformations. For canonical gravitational theories, the Poisson brackets
obeyed by the constraints show the way toward such a derivation, thanks to the
presence of a structure function closely related to the spatial
metric. Classically, the Poisson bracket of two Hamiltonian constraints is
given by
\begin{equation}
  \{H[N_1],H[N_2]\}= D[q^{ab}(N_1\partial_bN_2-N_2\partial_bN_1)]
\end{equation}
in Lorentzian signature. In spherical symmetry, this equation refers only to
the radial direction,
\begin{equation} \label{HH}
  \{H[N_1],H[N_2]\}= D[E^x(E^{\varphi})^{-2}(N_1N_2'-N_1'N_2)]\,.
\end{equation}
If we apply the previous canonical transformation, the structure function is
replaced by $4E^x/(\tilde{E}^{\varphi})^2$. Identifying this new coefficient
with the inverse radial metric according to the geometry of hypersurface
deformations, we obtain
$\tilde{q}_{xx}=\frac{1}{4} (\tilde{E}^{\varphi})^2/E^x=q_{xx}$. The
compatible space-time metric, derived from the structure function, is
therefore invariant under canonical transformations. This argument for a
spatial metric derived from a modified structure function has been made in
\cite{NonCovPol}, after an earlier more specific construction in
\cite{Absorb}. More recently, the first black-hole models derived from this
viewpoint were presented in \cite{SphSymmEff,SphSymmEff2,SphSymmEff3}. These
constructions used an additional ingredient \cite{EffLine} that determines the
transformation of lapse functions and shift vectors also using the
hypersurface-deformation brackets.

A similar derivation of a compatible metric is required for modifications that
are not simply canonical transformations of the classical theory. The
construction shows that several conditions must be realized for a compatible
space-time metric to exist. They must be fulfilled even if one uses
partial gauge fixings or deparameterization, sidestepping potential problems
with the complicated full gauge content of space-time theories, in a modified
theory. After all, a gauge fixing or preferred frame can be interpreted as a
slicing or coordinate choice in space-time only if one already knows that a
compatible space-time structure exists for a given modified theory.

Distilling the main conditions from the procedure briefly outlined for
our simple canonical transformation, we summarize the requirements for
consistent space-time models as follows:
\begin{itemize}
  \item We must have
a bracket of hypersurface-deformation form (\ref{HH}) in order to refer to the
structure function. For this to be possible, the modified constraints must,
first of all, remain first class, or be anomaly-free. The first-class condition is
necessary even in partially gauge-fixed or deparameterized theories, which
have often been argued to allow a relaxation of this condition. However, a
certain choice constitutes a gauge fixing of a modified theory only if the
theory remains gauge invariant. This condition cannot be checked at all if one
modifies a theory after a classical gauge has already been fixed. Similarly,
modifying a deparameterized theory does not guarantee that the
deparameterization fields constitute a valid choice of frame in a covariant theory.
\item Anomaly-free modifications of the Hamiltonian constraint must be such
  that the hypersurface-deformation form (\ref{HH}) is preserved. In
  particular, the Poisson bracket of two Hamiltonian constraints must be
  proportional to the diffeomorphism constraint. General covariance and
  space-time structure are off-shell properties that require a specific form
  of constraint brackets beyond the condition that they be closed.

  It is
    sometimes possible to work with phase-space dependent linear combinations
    of the constraints that are easier to handle from other perspectives, such
    as finding quantum representations. An example is given by partial
    Abelianizations as in \cite{LoopSchwarz,LoopSchwarz2}. However, the new
    constraints then are not identical with hypersurface-deformation
    generators and lack geometrical interpretations in space-time. If such
    linear combinations are used, any modification or quantization must be
    such that the existence of a linear transformation back to
    hypersurface-deformation brackets is preserved.
  \item Finally, if the first two conditions are fulfilled, there is an
    unambiguous modified structure function in a version of (\ref{HH}). For
    this structure function to determine a well-defined space-time structure,
    it must change under gauge transformations like an inverse spatial metric
    does under coordinate transformations, in order to be usable in a line
    element. This property is required to hold only on-shell, when the
    constraints and equations of motion are imposed. In fact, it can hold only
    on-shell because a comparison of canonical gauge transformations with
    coordinate changes requires a relationship between momenta and time
    derivatives of degrees of freedom. The latter is part of Hamilton's
    equations of motion in a canonical theory.  As shown in \cite{HigherCov},
    this new covariance condition in spherically symmetric models, when
    applied to the angular component $q_{\vartheta\vartheta}=E^x$ of the
    metric, requires that the Hamiltonian constraint does not depend on
    spatial derivatives of $K_x$, and that $\{\tilde{q}^{xx},H[\epsilon^0]\}$
    with a generic gauge function $\epsilon^0$ does not depend on spatial
    derivatives of $\epsilon^0$ on-shell.

    This important covariance condition is not implied by having anomaly-free
    constraint brackets of hypersurface-deformation form. It must therefore be
    checked independently and can be used to rule out modifications that would
    otherwise form a consistent deformation of the classical gauge theory. It
    follows that modifications of deparameterized models are not guaranteed to
    be covariant and have compatible space-time metrics for their solutions,
    even if they make use of an anomaly-free algebra. Deparameterization takes
    into account only the gauge structure of a canonical theory but not the
    relevant transformation properties of structure functions.
\end{itemize}
These restrictive conditions provide the foundations of emergent
modified gravity, and of covariant models of loop quantum gravity.

\subsection{General covariance and emergent modified gravity}
\label{s:EMG}

A compatible space-time line element that can be used for reliable geometrical
interpretations of a constrained canonical theory must be invariant not only
under canonical transformations but also under gauge transformations. The
Hamiltonian and diffeomorphism constraints, being first class, determine a
large set of gauge transformations for the canonical variables $(E^x,P_x;
E^{\varphi},P_{\varphi})$. The first-class nature of the constraints makes
sure that they are compatible with their own gauge transformations in the
sense that gauge-transformed solutions of the constraints obey the same
constraint equations. Since the dynamics is fully constrained, it
also respects the constraints.

However, this property of anomaly-freedom in general does not imply that
solutions of the theory allow a geometrical space-time interpretation in terms
of an invariant line element. For this to be the case, there must be some
correspondence between the canonical gauge transformations generated by the
constraints, which determine transformations for components of a candidate
space-time metric derived from the structure functions, and coordinate
transformations of a space-time metric tensor. The first-class nature of
modified constraints only implies that the number of gauge transformations is
not reduced by modifications, or that there are no gauge anomalies. General
covariance requires a stronger condition in order to make sure that the gauge
transformations are of a certain form compatible with changes of coordinates
or space-time slicings.

The full set of conditions is imposed in emergent modified gravity, where it
has been evaluated in spherically symmetric models on the same phase space as
in the classical theory, without additional variables and momenta that would
be implied by higher-derivative corrections. The same phase space has been
used in black-hole models of loop quantum gravity, which can therefore be
tested for covariance by a comparison with emergent modified gravity.

Here, we quote the main result from \cite{HigherCov}, giving a modified
Hamiltonian constraint for vacuum spherically symmetric gravity that is
strictly periodic in $P_{\varphi}$,
\begin{eqnarray}\label{H}
    H[N]
    &=& -\frac{1}{2G}\int{\rm d}x N \chi \sqrt{E^x} \Bigg[ E^\varphi \Bigg(
   -\frac{V}{2\sqrt{E^x}}
    + 2 \frac{\sin^2 \left(\bar{\lambda} P_\varphi\right)}{\bar{\lambda}^2}\left(\frac{\partial c_{f}}{\partial E^x}+\frac{c_f\alpha}{2E^x}\right)
    \nonumber\\
    &&+   4 \frac{\sin \left(2 \bar{\lambda} P_\varphi\right)}{2 \bar{\lambda}}
        \left(\frac{\partial q}{\partial E^x}+\frac{q\alpha}{2E^x}\right)
    \Bigg)
    + 4 P_x \left(c_f \frac{\sin (2 \bar{\lambda} P_\varphi)}{2 \bar{\lambda}}
    + q \cos(2 \bar{\lambda} P_\varphi)\right)
    \nonumber\\
    &&+ \frac{((E^x)')^2}{E^\varphi} \left(
    - \frac{\alpha}{4 E^x} \cos^2 \left( \bar{\lambda} P_\varphi \right)
    + \frac{P_x}{E^\varphi} \bar{\lambda}^2 \frac{\sin \left(2 \bar{\lambda} P_\varphi \right)}{2 \bar{\lambda}}
    \right)
    \nonumber\\
    &&
    + \left( \frac{(E^x)' (E^\varphi)'}{(E^\varphi)^2}
    - \frac{(E^x)''}{E^\varphi} \right) \cos^2 \left( \bar{\lambda} P_\varphi \right)
       \Bigg]\,.
\end{eqnarray}
The Poisson bracket of two of these constraints has a structure function
\begin{equation} \label{emergent}
    \tilde{q}^{x x}
    =
    \left(
    \left( c_{f}
    + \left(\frac{\bar{\lambda} (E^x)'}{2 E^\varphi} \right)^2 \right) \cos^2 \left(\bar{\lambda} P_\varphi\right)
    - 2 q \bar{\lambda}^2 \frac{\sin \left(2 \bar{\lambda} P_\varphi\right)}{2 \bar{\lambda}}\right)
    \chi^2 \frac{E^x}{(E^\varphi)^2}
\end{equation}
that can be used as inverse spatial metric in a compatible space-time line
element.  Up to canonical transformations and a sign choice (which would
replace trigonometric functions with hyperbolic functions), this constraint is
the most general modification of the classical expression with up to
second-order spatial derivatives of the phase-space variables.
  
The expression (\ref{H}) is more general than (\ref{Hmod}) not only in that
there are more modification functions, given by $\chi(E^x)$, $c_f(E^x)$,
$V(E^x)$, $\alpha(E^x)$ and $q(E^x)$ in addition to the constant parameter
$\bar{\lambda}$. It also has an extended derivative structure because terms
with spatial derivatives of $E^x$ and $E^{\varphi}$ may be multiplied by
modification functions depending on the momenta. (Or, equivalently, momentum
terms are multiplied by modification functions depending on spatial
derivatives of $E^x$ and $E^{\varphi}$.) The only modification functions that
do not change the structure function $\tilde{q}^{xx}$ are $V(E^x)$ and
$\alpha(E^x)$.  The former can be interpreted as the dilaton potential
(equalling $V(E^x)=-2/\sqrt{E^x}$ in the spherically symmetric reduction of
general relativity), which is a known modification of spherically symmetric
classical general relativity that does not change the derivative
structure. The latter modification function, $\alpha$, implies on-shell
modifications proportional to $(\dot{E}^x)^2-((E^x)')^2$ upon using the
equations of motion if all the other modification functions take their
classical values and the radial space-time is Minkowski. This modification is
therefore special in spherical symmetry where $E^x$ may be interpreted as a
scalar field on a $1+1$-dimensional space-time. The corresponding modification
is of Horndeski type in a $1+1$-dimensional scalar-tensor theory. All other
modification functions change the emergent metric and therefore have no analog
in higher-curvature actions.

While the restricted version (\ref{Hmod}) with independent modification
functions $F_2$, $\alpha$ and $\bar{\alpha}$ can be used to characterize most
models of loop quantum gravity that have been considered so far, as
demonstrated in Section~\ref{s:Types}, it does not implement the covariance
condition which requires that the structure function transforms as an inverse
spatial metric. The complete expression (\ref{H}) does include this condition
and therefore represents covariant Hamiltonian constraints. For this property,
including symmetry-restoring terms with spatial derivatives of $E^x$ and
$E^{\varphi}$ multiplying functions of the momentum $P_{\varphi}$ is
important. Such terms had not been included in (\ref{Hmod}) because the
classical derivative structure had been maintained in this case.

In \cite{HigherCov}, the constraint (\ref{H}) is embedded in a larger set of
modified constraints that appear different from (\ref{H}) but are strictly
related to it by canonical transformations. These properties are crucial for
an understanding of holonomy modifications in models of loop quantum
gravity. We will return to this question in Section~\ref{s:Schemes}.

As a special case of interest for holonomy modifications in models of loop
quantum gravity, if only $\bar{\lambda}$ takes a non-classical value we obtain
the constraint
\begin{eqnarray}\label{Hlambda}
    H[N]
    &=& -\frac{1}{2G}\int{\rm d}x N \sqrt{E^x} \Bigg[
   \frac{ E^\varphi }{E^x} \left(1+\frac{\sin^2 \left(\bar{\lambda} P_\varphi\right)}{\bar{\lambda}^2}\right)
    + 4 P_x \frac{\sin (2 \bar{\lambda} P_\varphi)}{2 \bar{\lambda}}
     \nonumber\\
    &&+ \frac{((E^x)')^2}{E^\varphi} \left(
    - \frac{\cos^2 \left( \bar{\lambda} P_\varphi \right)}{4 E^x} 
    + \frac{P_x}{E^\varphi} \bar{\lambda}^2 \frac{\sin \left(2 \bar{\lambda} P_\varphi \right)}{2 \bar{\lambda}}
    \right)
    \nonumber\\
    &&
    + \left( \frac{(E^x)' (E^\varphi)'}{(E^\varphi)^2}
    - \frac{(E^x)''}{E^\varphi} \right) \cos^2 \left( \bar{\lambda} P_\varphi \right)
       \Bigg]
\end{eqnarray}
and the structure function
\begin{equation} \label{emergentlambda}
    \tilde{q}^{x x}
    =
    \left(1
    + \left(\frac{\bar{\lambda} (E^x)'}{2 E^\varphi} \right)^2 \right) \cos^2
  \left(\bar{\lambda} P_\varphi\right) \frac{E^x}{(E^\varphi)^2}
    \,.
\end{equation}
This example is instructive for the importance of the emergent metric: If we
solve the constraints and equations of motion in the Schwarzschild gauge,
where $P_x=0$ and $P_{\varphi}=0$ as static gauge conditions, the reduced
Hamiltonian constraint and the equations of motion take the classical
form. The solution for $(E^{\varphi})^2/E^x$ is therefore classical, which
would suggest an unmodified space-time solution if the emergent metric is
ignored. In other gauges, however, such as Gullstrand--Painlev\'e,
$P_{\varphi}$ is non-zero and implies modified solutions. This mismatch shows
that there is no coordinate transformation between the resulting line
elements, and the theory is not covariant. The problem is solved by using the
emergent metric (\ref{emergentlambda}), which contains a non-classical
$\bar{\lambda}$-term even if $P_{\varphi}=0$.

An explicit demonstration that the resulting Schwarzschild and
Gullstrand--Painlev\'e gauges are related by a coordinate transformation of
their emergent space-time line elements can be found in
\cite{SphSymmEff2}. Our discussion here provides general context for a
demonstration of non-covariance made in the specific model of
\cite{LoopGauge1}, using the same slicings. In this case, modifications had
been implemented after fixing the gauge, which prevents a systematic analysis
of covariance properties. (Nevertheless, it had already been noted in
\cite{LoopGauge1} itself that solutions of the model in different gauges are
not related by co-ordinate transformations.) In particular, deriving the
compatible emergent metric requires full access to all space-time gauge
transformations. Emergent modified gravity shows that a modification as used
in \cite{LoopGauge1} can be made covariant by combining the symmetry-restoring
terms in (\ref{Hlambda}) with the canonical transformation (\ref{canonicalx})
discussed later in order to introduce a non-constant $\lambda(E^x)$ replacing
$\bar{\lambda}$, provided the emergent space-time metric is used.

\section{Implications for modified spherically symmetric gravity}
\label{s:Implications}

The conditions summarized in Section~\ref{s:Types} include anomaly-freedom,
which is a prerequisite for general covariance and the existence of consistent
geometrical space-time interpretations of solutions of canonical theories. In
addition, the resulting candidate radial metric component, derived from the
modified structure function $|\tilde{q}^{xx}|$, must transform under gauge
transformations as a radial metric component under coordinate changes. Only
this condition makes it possible to interpret
\begin{equation}
  {\rm d}s^2=-\sigma N^2{\rm d}t^2+|\tilde{q}_{xx}| ({\rm d}x+M{\rm d}t)^2+E^x({\rm
    d}\vartheta^2+\sin^2(\vartheta) {\rm d}\varphi^2)
\end{equation}
as an invariant space-time line element. Here,
$\sigma={\rm sgn}(\tilde{q}^{xx})$ determines the space-time signature and has
to be included because the structure function $\tilde{q}^{xx}$ is not always
guaranteed to be strictly positive. For more details about signature change in
emergent modified gravity we refer to \cite{HigherCov,EmergentSig}.

\subsection{Lack of covariance and symmetry-restoring terms in models of
  loop quantum gravity}

For an anomaly-free theory of the form (\ref{Hmod}), the absolute value of the
inverse structure function takes the form
\begin{equation}
  \frac{(E^{\varphi})^2}{|\beta|E^x}=
  \frac{1}{|\bar{\alpha}\bar{\alpha}_{\Gamma} \partial F_2/\partial 
    P_{\varphi}|} \frac{(E^{\varphi})^2}{E^x}
\end{equation}
using (\ref{beta}). However, this function does not obey the covariance
condition, as shown by the results of \cite{HigherCov}. Anomaly-freedom is
therefore a necessary but not a sufficient condition for general covariance
and the existence of a compatible space-time metric.

The covariance condition on the structure function has been evaluated in full
generality in emergent modified gravity. As a result, none of the models
originally suggested in loop quantum gravity, which are all of the form
(\ref{Hmod}), are covariant. However, they can be made covariant by additional
modifications that include symmetry-restoring terms, as first constructed and
analyzed in \cite{SphSymmEff,SphSymmEff2}. Their form depends on the canonical
variables used. In (\ref{H}), for instance, contributions that are periodic
functions of $P_{\varphi}$ multiplied by spatial derivatives of the densitized
triad are new compared with traditional models of loop quantum gravity. From
the viewpoint of this theory, such terms are unexpected because the classical
derivative terms are completely independent of $P_{\varphi}$. A justification
from loop quantum gravity would therefore require an involved choice of
holonomies in combination with fluxes evaluated at different points on the
radial line.

These terms may look different after a canonical transformation, which might
help to suggest alternative motivations for this kind of modified gravity. For
instance, the periodic function can be replaced by square roots of quadratic
terms in $P_{\varphi}$ if a canonical transformation
\begin{equation}
 \frac{ \sin(\bar{\lambda}P_{\varphi})}{\bar{\lambda}}\mapsto P_{\varphi}\quad,\quad
 E^{\varphi}\mapsto \frac{E^{\varphi}}{\cos(\bar{\lambda}P_{\varphi})}
\end{equation}
is applied, while $(P_x,E^x)$ remains unchanged. Any single cosine factor in
$\sin(2\bar{\lambda}P_{\varphi})=2\sin(\bar{\lambda}P_{\varphi})\cos(\bar{\lambda}P_{\varphi})$
is then mapped to a square root $\sqrt{1-\bar{\lambda}^2P_{\varphi}^2}$,
featuring a new characteristic dependence of the covariant Hamiltonian
constraint on $P_{\varphi}$. The same transformation introduces spatial
first-order derivatives of the new $P_{\varphi}$ through the original spatial
derivatives of $E^{\varphi}$. It is therefore possible to modify spherically
symmetric general relativity by introducing spatial derivatives of the
momenta, while keeping the theory compatible with general covariance. However,
the resulting modified constraints arising from the addition of spatial
derivatives of the momenta will always be related to (\ref{H}) by a suitable
canonical transformation. Physical implications of such related theories are
always equivalent, but one of them may be preferable for finding possible
relationships to specific quantization steps.

\subsection{Lattice refinement and holonomy schemes}
\label{s:Schemes}

Models of loop quantum gravity often consider different rules for possible
$E^x$-dependent modifications of the momentum terms, as in a function of the
form $\sin(\delta P_{\varphi}/\sqrt{E^x})$. Specific choices are then
motivated by phenomenological considerations, such as the low-curvature or
near-horizon behavior of space-time. In general, a decreasing argument
$P_{\varphi}/\sqrt{E^x}$ with increasing $E^x$ is preferred because
$P_{\varphi}$ is not necessarily small at large distances from a black-hole
model in particular in the example of a Schwarzschild-de Sitter black hole. In
this case, the classical solution suggests that $P_{\varphi}\propto \dot{a}$
grows at late times like the proper-time derivative of the scale factor in de
Sitter space-time. The ratio $P_{\varphi}/\sqrt{E^x}$, by contrast, is then
proportional to the constant Hubble parameter and may be small if
$\delta\Lambda$ is small for a given choice of $\delta$ in holonomy
modifications and the cosmological constant $\Lambda$.

\subsubsection{Heuristics}

An additional factor of $1/\sqrt{E^x}$ in holonomy arguments, combined to a
holonomy function $\lambda(E^x)=\delta/\sqrt{E^x}$ as the multiplier of
$P_{\varphi}$, can be interpreted as an implication of lattice refinement
\cite{InhomLattice}, in which an underlying discrete lattice structure of
space is assumed. Increasing $E^x$, giving the area of symmetric spheres in
spherically symmetric models, would magnify any constant coordinate area
$\delta^2$ of an individual lattice plaquette with constant coordinate side
length $\bar{\lambda}=\delta$, which has the metric area $\delta^2 E^x$. By
dividing the side length by $\sqrt{E^x}$, refining the lattice at larger
values of $E^x$, then cancels out the metric factor and implies a constant
metric area $E^x (\delta/\sqrt{E^x})^2=\delta^2$. Depending on how the lattice
gets refined, other powers of $E^x$ may also appear. In models of loop quantum
gravity, such effects are often simplified to a dichotomy between a
``$\mu_0$-scheme'' (constant holonomy function) and a
``$\bar{\mu}$-scheme'' (decreasing holonomy function $\delta/\sqrt{E^x}$). 

The traditional derivation of such coefficients and the distinction between
$\bar{\mu}$ and $\mu_0$ aims to relate length parameters of holonomies, such
as $\lambda$ in $\sin(\lambda P_{\varphi})$, to the smallest non-zero area
eigenvalue $\Delta$ in loop quantum gravity. Equating $m \lambda^2$ with
$\Delta$, using a suitable metric coefficient $m$, then suggests that spatial
discreteness of an underlying quantum space-time is the origin of modified
dynamics. The value of $\Delta$ determines the smallest possible area of a
square loop along whose edge one can compute a holonomy. The ambiguity of
different holonomy schemes appears because the metric factor $m$ may be chosen
in different ways. This factor is phase-space independent, resulting in a
constant $\lambda$, if the coordinate area is equated to $\Delta$, and it depends
on $E^x$ as $\lambda\propto \sqrt{\Delta/E^x}$ if the area is computed using the
classical spatial metric on a portion of a symmetric sphere. The latter is
often argued to be physically more meaningful, for instance in \cite{APSII},
because it uses metric areas rather than coordinate areas.

However, these arguments are incomplete for several reasons. First, the
relevant term in the full Hamiltonian constraint in which these holonomies
appear rewrites the Yang-Mills-type curvature using a version of the
non-Abelian Stokes theorem; see for instance \cite{RS:Ham,AnoFree}. This theorem is
topological and does not require a metric. It is therefore unclear how a
metric or the area spectrum should enter these expressions. Secondly, the
classical phase-space form of the metric is not necessarily preserved after
modifications. A covariance analysis is then required in order to determine
the correct emergent metric, but this step has not been included in
traditional discussions of holonomy schemes.  Since the mathematical
appearance of regimes in which lattice-refinement arguments are invoked
depends on choices of slicings or coordinates, it is important to perform such
an analysis in a generally covariant setting of effective equations that have
the required symmetries. Restricting considerations to a fixed slicing or a
single deparameterization of the canonical theory with an internal time is
inconclusive. Finally, the defining features of holonomy schemes are not
invariant under canonical transformations, and therefore cannot imply
physically distinct effects as claimed in the context of holonomy schemes
going back to \cite{APSII}.

\subsubsection{A universal scheme for holonomy modifications}
\label{s:Universal}

In a covariant theory, an $E^x$-dependence of holonomy functions is possible
according to emergent modified gravity, but with two additional properties
that have rarely been considered: Such a dependence is incompatible with
strictly periodic functions of $P_{\varphi}$, and it is always canonically
equivalent to a modified constraint with $E^x$-independent modification of the
momentum terms \cite{HigherCov} but different functional forms of other
modification functions. Here, we present a detailed comparison between
different holonomy schemes in a universal setting that considers covariance as
well as equivalence classes of canonically related formulations of the same
physical theory. The first property can be seen also in \cite{Absorb} and the
recent \cite{MassCovariance}, but the freedom of performing canonical
transformations had not been analyzed in these papers. A reliable analysis of
potential physical implications of holonomy-modifications schemes can only be
obtained if relevant properties are independent of the choice of canonical
variables. Our discussion here fills this important gap.

In modified $1+1$-dimensional models with up to second-order spatial
derivatives of the canonical variables, as usually considered in models of
loop quantum gravity, any covariant Hamiltonian constraint is equivalent to an
expression with a constant holonomy function $\bar{\lambda}$, given by
(\ref{H}). This result can already be seen by the restricted example of
(\ref{momentumterms}), where the momentum-dependent terms
$\bar{P}_{\varphi}=P_{\varphi}/\sqrt{E^x}$ in the periodic functions and
\begin{equation}
  \label{canonicalPP}
  \bar{P}_x=E^xP_x-\frac{1}{2}E^{\varphi}P_{\varphi}
\end{equation}
in one of the coefficients are simply momenta of
$\bar{E}^{\varphi}=\sqrt{E^x}E^{\varphi}$ and $\bar{E}^x=E^x$ after a combined
canonical transformation.  Considering the freedom of applying canonical
transformations, therefore, there is no difference between a lattice-refining
(or $\bar{\mu}$-type) modification with periodic functions depending on
$P_{\varphi}/\sqrt{E^x}$, and a ($\mu_0$-type) modification with constant
coefficients in periodic functions.  On a formal level, there is no physical
distinction between the two choices (or others) of canonical variables because
$(E^x,E^{\varphi})$ are as good as phase-space parameterizations of the
spatial geometry as $(\bar{E}^x,\bar{E}^{\varphi})$. While only the former
directly corresponds to the components of a classical densitized triad, none
of them are densitized-triad components of the spatial part of the emergent
line element. Both choices are therefore equally admissible in a modified
constraint.

In the full setting of emergent modified gravity, holonomy-type modifications
are related to the constant parameter $\bar{\lambda}$, which appears non-trivially in
the emergent metric. The appearance of holonomy-like terms depends on the choice of
canonical phase-space coordinates. Generalizing the preceding example, it is possible to apply a
canonical transformation
\begin{eqnarray}
  P_{\varphi}=\frac{\lambda}{\bar{\lambda}}\tilde{P}_{\varphi}\quad&,&\quad
  E^{\varphi}=\frac{\bar{\lambda}}{\lambda} \tilde{E}^{\varphi}\\
  P_x=\tilde{P}_x+\tilde{E}^{\varphi}\tilde{P}_{\varphi}\frac{{\rm
  d}\ln\lambda}{{\rm d}\tilde{E}^x} 
  \quad&,&\quad E^x=\tilde{E}^x \label{canonicalx}
\end{eqnarray}
where $\lambda$ is an arbitrary function of $\tilde{E}^x$. The constant
$\bar{\lambda}$ in the periodic dependence of the Hamiltonian constraint on
$\bar{\lambda}P_{\varphi}$ is then replaced by a scale-dependent function
$\lambda(\tilde{E}^x)$, and the Hamiltonian constraint (\ref{Hlambda}) takes
the form
\begin{eqnarray}\label{Hlambda2}
    &&H[N]
    = -\frac{1}{2G}\int{\rm d}x N \sqrt{\tilde{E}^x}  \frac{\lambda}{\bar{\lambda}}\Bigg[
   \frac{ \tilde{E}^{\varphi}}{\tilde{E}^x} \left(\frac{\bar{\lambda}^2}{\lambda^2}  +\frac{\sin^2 (\lambda
        \tilde{P}_\varphi)}{\lambda^2}\right) \nonumber\\
&&    + \frac{4}{\tilde{E}^x}\left( \tilde{E}^x\tilde{P}_x +
   \tilde{E}^{\varphi}\tilde{P}_{\varphi}\frac{{\rm
  d}\ln\lambda}{{\rm d}\ln\tilde{E}^x} \right)
   \frac{\sin (2 \lambda \tilde{P}_\varphi)}{2\lambda}
     \nonumber\\
    &&+  \frac{((\tilde{E}^x)')^2}{(\tilde{E}^\varphi)^2} \left(
       \frac{\tilde{E}^{\varphi}}{\tilde{E}^x}  \left(\frac{{\rm
       d}\ln\lambda}{{\rm d}\ln\tilde{E}^x} -\frac{1}{4}\right) \cos^2 ( \lambda
       \tilde{P}_\varphi) 
    + \frac{\lambda^2}{\tilde{E}^x}\left(\tilde{E}^x\tilde{P}_x+
       \tilde{E}^{\varphi}\tilde{P}_{\varphi} \frac{{\rm 
  d}\ln\lambda}{{\rm d}\ln\tilde{E}^x}  \right)  \frac{\sin(2
       \lambda \tilde{P}_\varphi)}{2\lambda}
    \right)
    \nonumber\\
    &&
    + \left( \frac{(\tilde{E}^x)'
       (\tilde{E}^\varphi)'}{(\tilde{E}^\varphi)^2}
    - \frac{(\tilde{E}^x)''}{\tilde{E}^\varphi} \right) \cos^2 ( \lambda
       \tilde{P}_\varphi ) 
       \Bigg]\,.
\end{eqnarray}
Various occurrences of ${\rm d}\ln\lambda/{\rm d}\ln\tilde{E}^x$ directly
determine the power-law exponent of $\lambda(\tilde{E}^x)$ in a possibly
phenomenological parameterization of the form
$\lambda\propto (\tilde{E}^x)^q$.  However, unless $q=0$, any transformation
of this form introduces a term that is non-periodic in $\tilde{P}_{\varphi}$
via the second contribution to $P_x$ in (\ref{canonicalx}). Since (\ref{H}) is
the most general covariant Hamiltonian with the given derivative order, it
follows that a strictly periodic dependence on $P_{\varphi}$, as suggested by
holonomy modifications, is possible only with a scale-independent coefficient
$\bar{\lambda}$.

General covariance therefore imposes strict new conditions on possible
modifications that are not compatible with common assumptions about holonomy
modifications in models of loop quantum gravity. These restrictions cannot be
seen in isotropic cosmological models, in which holonomy schemes had been
formulated first but where spatial homogeneity trivializes covariance
conditions. In this context, realized within spherically symmetric models by
the asymptotic region of Schwarzschild-de Sitter black holes, $E^{\varphi}$ as
well as $E^x$ are proportional to the squared scale factor, while
$P_{\varphi}$ is proportional to its proper-time derivative. In the
traditional $\mu_0$-scheme, a constant $\bar{\lambda}$ is used in holonomy
modifications depending on $\bar{\lambda}P_{\varphi}$. Since the proper-time
derivative $\dot{a}$ of the scale factor grows rapidly in the presence of a
cosmological constant, holonomy modifications are large even in semiclassical
regimes. The $\bar{\mu}$-scheme proposes to use instead a decreasing function
$\lambda\propto 1/\sqrt{E^x}\propto 1/a$, such that $\lambda \dot{a}$ is
proportional to the Hubble parameter, which is constant in a semiclassical
regime dominated by a cosmological constant. Holonomy modifications can then
maintain a small value for a long time.

In the present context, the Hubble parameter can be used as a new momentum
$\tilde{P}_{\varphi}$ obtained from the original $P_{\varphi}$ from
(\ref{canonicalx}) with the specific functional dependence of $\lambda$ as
used in the $\bar{\mu}$-scheme of isotropic cosmological models. The same
transformation replaces the original $E^{\varphi}$, proportional to the
squared scale factor, with an expression proportional to
$\sqrt{\tilde{E^x}}\tilde{E}^{\varphi}$, which is proportional to the volume of
a spatial reference region. In the canonical pair
$(E^{\varphi},P_{\varphi})\propto (a^2,\dot{a})$, a non-constant $\lambda$
would have to be used for small holonomy modifications at late times, while
the pair $(\tilde{E}^{\varphi},\tilde{P}_{\varphi})\propto(a^3,\dot{a}/a)$
implies small late-time holonomy modifications even for constant
$\bar{\lambda}$.  The canonical transformation (\ref{canonicalx}) can be seen
as a systematic generalization of the change of canonical variables proposed
in \cite{APSII} for isotropic models.

However, in contrast to the traditional distinction between $\bar{\mu}$ and
$\mu_0$ schemes, covariant Hamiltonian constraints based on the sets
$(E^{\varphi},P_{\varphi})$ and $(\tilde{E}^{\varphi},\tilde{P}_{\varphi})$
are not physically distinct. If one starts with (\ref{Hlambda}), in which only
a constant modification parameter $\bar{\lambda}$ appears, the transformed
version (\ref{Hlambda2}) not only has a non-constant $\lambda$ in holonomy
terms given by trigonometric functions, this function also appears in several
other places that, compared with the general expression (\ref{H}), correspond
to non-constant choices for the function $\chi$, $V$, $c_f$, and $\alpha$. The
full set of modification functions allowed within a covariant setting should
therefore be considered when a holonomy-modification scheme is discussed. If a
canonical transformation is applied to the phase-space variables, it cannot be
expressed merely as a change of the holonomy function $\lambda$. It rather
implies a complicated transformation of the full set of modification functions.

With these insights, the question remains as to how physically distinct
implications of modification functions can be seen within the complete set of
canonically equivalent theories. As first discussed in \cite{EmergentMubar},
the traditional distinction between different schemes of holonomy
modifications can only be made on the level of equations of motion, rather
than directly in the off-shell Hamiltonian constraint. It is then possible to
suppress holonomy effects in semiclassical regimes by suitable choices of the
full set of modification functions. For instance, (\ref{Hlambda2}) is
equivalent to (\ref{Hlambda}) and therefore implies large holonomy
modifications in some semiclassical regimes (large $\tilde{E}^x$) even if
it uses a decreasing holonomy function $\lambda(\tilde{E}^x)$.

The reason for this behavior cannot be found in the holonomy terms but rather
in the overall multiplier $\lambda/\bar{\lambda}$ in (\ref{Hlambda2})
introduced by the canonical transformation. This function can be interpreted
as a decreasing choice of $\chi$ in the general expression (\ref{H}). Its
implications can be seen in Hamilton's equation for
\begin{equation} \label{Exdot}
  \dot{\tilde{E}}{}^x\propto \chi c_f\sqrt{\tilde{E}^x}
  \frac{\sin(2\lambda \tilde{P}_{\varphi})}{2\lambda}
  \left(1+\lambda^2\frac{((\tilde{E}^x)')^2}{4(\tilde{E}^{\varphi})^2}\right) 
\end{equation}
implied by (\ref{Hlambda2}) with some coefficients interpreted as the original
modification functions $\chi$ and $c_f$ in (\ref{H}).  The phenomenological
aim of lattice-refinement effects is to make sure that holonomy modifications
are not large in regimes that are expected to be semiclassical, such as the
large-$\tilde{E}^x$ region of a Schwarzschild-de Sitter space-time in which
$\tilde{P}_{\varphi}$ would be non-zero in Schwarzschild-type coordinates
while $(\tilde{E}^x)'/\tilde{E}^{\varphi}$ remains constant. For decreasing
$\lambda(\tilde{E}^x)$, the parenthesis in (\ref{Exdot}) can therefore be
ignored, and one would like $\sin(2\lambda \tilde{P}_{\varphi})/(2\lambda)$ to
be close to the classical
$\tilde{P}_{\varphi}\propto \dot{\tilde{E}}{}^x/\sqrt{\tilde{E}^x}$ for nearly
classical behavior. This condition requires that
\begin{equation}
  \frac{\sin(2\lambda \tilde{P}_{\varphi})}{2\lambda} \sim
  \frac{\dot{\tilde{E}}{}^x}{\sqrt{\tilde{E}^x}}
\end{equation}
which is then compatible with nearly classical solutions for $\tilde{E}^x$ and
$\tilde{P}_{\varphi}$. However, equation (\ref{Exdot}) implies an
additional factor of $1/\chi=\bar{\lambda}/\lambda$  in the expression for $\sin(2\lambda
\tilde{P}_{\varphi})/(2\lambda)$ according to (\ref{Hlambda2}), which is an increasing function of
$\tilde{E}^x$ and therefore prevents $\sin(2\lambda
\tilde{P}_{\varphi})/(2\lambda)$ from being small for near-classical
solutions.

This discussion demonstrates how the large-scale behavior of (\ref{Hlambda})
is maintained after a canonical transformation (\ref{canonicalx}), even though
it might appear that small holonomy modifications are implied by a decreasing
holonomy function $\lambda(\tilde{E}^x)$. As already mentioned, the physical
behavior of a modified theory  within an equivalence class of canonically
related formulations is determined not by a single modification function but
by the full set of possible modifications. A decreasing $\chi$ then
counteracts the effects of a decreasing $\lambda$. Conversely, one could stay
in the original canonical theory (\ref{H}) with a constant $\bar{\lambda}$ and introduce
an increasing modification function $\chi(E^x)=\bar{\lambda}/\lambda(E^x)$ with a
desired form for $\lambda(E^x)$. The equation of motion
\begin{equation} \label{Exdot2}
  \dot{E}{}^x\propto \chi \sqrt{E^x}
  \frac{\sin(2\bar{\lambda} P_{\varphi})}{2\bar{\lambda}}
  \left(1+\bar{\lambda}^2\frac{((E^x)')^2}{4(E^{\varphi})^2}\right) 
\end{equation}
in these variables now shows that
\begin{equation}
  \frac{\sin(2\bar{\lambda} P_{\varphi})}{2\bar{\lambda}} \propto
  \frac{1}{\chi} \frac{\dot{E}^x}{\sqrt{E^x}}
\end{equation}
contains a decreasing factor $1/\chi$ that dynamically leads to small holonomy
modifications in semiclassical regimes. 

\subsubsection{Summary}

Comparing this picture with the traditional arguments for different holonomy
schemes, we notice important differences.  Traditionally, it is argued that
one has to construct a suitable function $\lambda(E^x)$ such that
$\lambda(E^x)P_{\varphi}$ is small in semiclassical regimes even while
$\bar{\lambda}P_{\varphi}$ remains of the order one or larger.  In order to
find a function $\lambda(E^x)$ that obeys this condition, phenomenological
discussions of this type make use of the classical $P_{\varphi}$ to determine
at which rate $\lambda(E^x)$ should decrease such that
$\lambda(E^x)P_{\varphi}$ remains small. However, in the resulting modified
theory, the function $\sin(\lambda(E^x)P_{\varphi})$ is then subject to
modified equations of motion, which in general are not compatible with the
classical behavior of $P_{\varphi}$ that was the starting point of such
discussions. The arguments are therefore intrinsically inconsistent. Moreover,
heuristic rules for explicit constructions of $\lambda(E^x)$ often require a
spatial metric, which is assumed to have the classical components
$(E^{\varphi})^2/E^x$ and $E^x$ for the radial and angular parts,
respectively. However, the resulting modifications are no longer compatible
with the classical line element and instead require an emergent space-time
metric.

A suitable choice of modification functions should  be based on the
equations of motion of the modified theory, such as (\ref{Exdot}), rather than
an individual holonomy term. Instead of
trying to find a function $\lambda(E^x)$ that replaces the constant
$\bar{\lambda}$ and is sufficiently small in semiclassical regimes, suppressing
holonomy modifications, one might as well choose a function $\chi$ (instead of
the classical constant $\chi=1$) such that $\dot{E}^x/(\chi \sqrt{E^x})$ and
therefore $\sin(2\bar{\lambda}P_{\varphi})/(2\bar{\lambda})$ becomes
sufficiently small. A modified $\chi$ is indeed implied if we start with a
non-constant $\lambda$ and then apply the inverse of (\ref{canonicalx}) in
order to obtain a constant holonomy length $\bar{\lambda}$. The same
transformation also implies a non-classical function $c_f$ as another
contribution to modified equations of motion: The derivative of
$\ln\lambda$ in this equation can then be included in a new $c_f$-term. Both
functions appear in the equation of motion relating $\dot{E}^x$ to $P_{\varphi}$.

In this way, lattice-refinement effects are compatible with constant holonomy
length, and $\mu_0$-type and $\bar{\mu}$-type schemes are just different
choices of canonical variables if they are combined with suitable modification
functions. This result is important for loop quantum gravity because only
constant holonomy length ($\mu_0$-type schemes) can be implemented by a
Hamiltonian constraint quantized in terms of basic holonomy-flux operators in
a closed commutator algebra; see \cite{EmergentMubar}. Canonical relationships
to effective constraints with non-constant holonomy length $\lambda(E^x)$ can
then be used to facilitate studies of the semiclassical limit because they
allow Taylor expansions of relevant trigonometric functions. In this way,
emergent modified gravity has reconciled competing notions of holonomy schemes
by combining their advantages within a single consistent setting of covariant
equivalence classes of canonically related formulations.

\subsection{Beyond vacuum spherical symmetry}

Emergent modified gravity has been extended from vacuum spherical symmetry to
the inclusion of scalar \cite{EmergentScalar} and perfect-fluid matter
\cite{EmergentFluid}, as well as an electromagnetic field \cite{EmergentEM}. In both
cases, non-trivial covariance conditions on the matter contributions have to
be evaluated, but they do not significantly restrict the types of different
modification functions as described for vacuum spherical symmetry. Similarly,
an extension to polarized Gowdy models \cite{EmergentGowdy} shows the same
behavior if spherical symmetry is relaxed. In particular, scalar matter and
polarized Gowdy models allow covariant modifications of holonomy type for
models with local physical degrees of freedom, in contrast to earlier attempts
without an emergent metric \cite{GowdyCov}.

While polarized Gowdy models do not significantly restrict the types of
modification functions, there are additional results about the form of
holonomy modifications as described in more detail in
\cite{EmergentGowdy}. The main difference is related to the extension of the
underlying phase space, where the single angular momentum component
$P_{\varphi}$ is replaced by two independent momenta, $P_X$ and $P_Y$, for the
geometry of possibly anisotropic planar wave fronts. Classically, these two
momenta are canonically conjugate to densitized-triad components $E^X$ and
$E^Y$ that appear in the line element \cite{EinsteinRosenAsh}
\begin{equation}
  {\rm d}s^2= -N^2{\rm d}t^2+\frac{E^XE^Y}{\epsilon} ({\rm
    d}\theta+N^{\theta}{\rm d}t)^2+ \epsilon\left(\frac{E^Y}{E^X}{\rm
    d}X^2+\frac{E^X}{E^Y}{\rm d}Y^2\right)\,.
\end{equation}

Holonomy modifications can be formulated with the same freedom of
choosing a constant $\bar{\lambda}$, or a function $\lambda$ that may depend
on the planar area $\epsilon$ (which is analogous to $E^x$ in spherically
symmetric models) and also on $\bar{W}=\ln\sqrt{E^Y/E^X}$ (which is a new anisotropy
degree of freedom compared with spherically symmetric models). In any
covariant Hamiltonian constraint, the choices of $\bar{\lambda}$ and
$\lambda$, respectively, are related by canonical transformations as described
for spherically symmetric models.  However, covariant holonomy modifications
in polarized Gowdy models require an additional triad dependence because
$\bar{\lambda}$ or $\lambda$ does not multiply a single momentum component,
but rather the specific combination
\begin{equation}
  P=\frac{E^XP_X+E^YP_Y}{\sqrt{E^XE^Y}}=e^{-\bar{W}}P_X+e^{\bar{W}}P_Y\,.
\end{equation}
Even for constant $\bar{\lambda}$, covariant holonomy modifications must
depend on components of the densitized triad, precluding the application of
basic holonomy operators in potential quantizations. Unlike scale-dependencies
of the form  $\lambda(E^x)$ in spherically symmetric models or
$\lambda(\epsilon)$ in polarized Gowdy models, the required dependence of $P$
is on the anisotropy degree of freedom through $\exp(\pm \bar{W})$. A possible
relationship with lattice refinement, which postulates that the holonomy
length decreases as a symmetry orbit expands but does not suggest a specific
dependence on anisotropies, is therefore harder to justify.

\subsection{Carrollian limits}

The analysis underlying emergent modified gravity has shown that it is
possible to obtain inequivalent space-time structures from phase-space
dependent modification functions in the Hamiltonian constraint. As a simple
example, it is possible to replace the classical $H[N]$ with
$H[\gamma_1 N]+D[\gamma_2 N]$ where $\gamma_1$ and $\gamma_2$ are suitable
phase-space dependent functions, and derive non-Lorentzian emergent
4-dimensional geometries. This result has led to new types of non-singular
signature change \cite{EmergentSig} where the emergent space-time turns
Euclidean in some regions, and it is possible to construct theories whose
emergent 4-dimensional geometry has the structure of the Carrollian limit of
space-time as defined in \cite{Carrollian,CarrollianADM}. The Carrollian limit
has a vanishing structure function, which can be evaluated in emergent
modified gravity by setting $\tilde{q}^{xx}=0$ instead of imposing the
covariance condition.

Carrollian versions of the gravitational constraints have been constructed for
different purposes in models of loop quantum gravity. In
\cite{LoopSchwarz,LoopSchwarz2} for spherical symmetry and
\cite{GowdyCov} for Gowdy configurations, such models are referred to as partial
Abelianizations because they lead to commuting Hamiltonian constraints. The
absence of any structure function in this case implies simplifications for
possible quantizations. In \cite{LoopShock}, a similar model has been obtained
from gauge fixings.

However, the incomplete status of such models in terms of covariance has
already been pointed out in \cite{SphSymmCov,GowdyCov}, and it follows again from
emergent modified gravity because such models are not of the general form
allowed by the covariance condition for Lorentzian geometries. These models
construct only modified or quantum versions of the Carrollian limit of
gravity, but not of full gravity which includes a complete Lorentzian regime.

\subsection{Well-posedness}

The result that emergent modified gravity is able to encompass Lorentzian and
Euclidean signature in the same covariant setting requires care in choosing
well-posed initial or boundary value problems. Early versions of models of
loop quantum gravity, which did not fully incorporate covariance, already
pointed this out, for instance in \cite{Action,SigChange,SigImpl}. If the sign
of the structure function changes, the only compatible candidate 4-dimensional
geometry has Euclidean signature in the radial space-time and therefore
requires boundary rather than initial-value problems for radial motion. The
angular part of the space-time metric is not determined by structure functions
in a spherically symmetric model and may be chosen with positive or negative
signs. The full space-time signature with negative structure function may
therefore be $(----)$ or $(--++)$. Continuity conditions across the
signature-change hypersurfaces can be used to choose a preferred version
\cite{Emergentq}.

It is impossible to view the corresponding dynamical equations as evolution of
initial values in the radial direction because there simply is no Lorentzian
4-dimensional geometry in this case. Claims that such equations have dynamical
instabilities and can be used to rule out certain models by observations
\cite{QCExclude} are therefore incorrect. (Results compatible with
observations are obtained if a well-posed initial-boundary value problem is
used as discussed in \cite{SigImpl}; see for instance \cite{SigChangeCMB} for
an explicit demonstration.)

\subsection{Time-reversal symmetry}
\label{s:TimeReversal}

The classic analysis \cite{Regained} of general relativity in canonical form
assumed that the spatial metric is one of the basic phase-space variables and
concluded that second-order equations of motion must be equivalent to those of
general relativity with only two free parameters, given by Newton's constant
and the cosmological constant. Working with an emergent metric that cannot be
turned into a phase-space variable by a local canonical transformation, for
instance because it contains spatial derivatives, makes it possible to evade this
conclusion and obtain modified dynamics with second-order time
derivatives.

Another assumption in \cite{Regained} was time-reversal symmetry, which can
also be relaxed in emergent modified gravity if the modification function $q$
in (\ref{H}) is non-zero. Properties and physical implications are discussed
in \cite{Emergentq}. Here, we only mention that it is possible to exploit this
modification function in order to obtain modified dynamics in which the
classical radial metric is compatible with gauge transformations, presenting
another example in which the uniqueness results from \cite{Regained} can be
circumvented.

\subsubsection{Classical structure function}

Such an example can also be found as the first modification of
\cite{MassCovariance}, which makes use of a complex Hamiltonian constraint
\begin{eqnarray}\label{HComplex}
    H[N]
    &=& -\frac{1}{2G}\int{\rm d}x N \sqrt{E^x}  \Bigg[
   \frac{ E^{\varphi}}{E^x} \left(1  +3\frac{\sin^2 (\lambda
        \tilde{P}_\varphi)}{\lambda^2}\right) \nonumber\\
&&    + \frac{4}{E^x}\left( E^xP_x -\frac{1}{2}
   E^{\varphi}P_{\varphi} \right)
   \frac{\sin (2 \lambda \tilde{P}_\varphi)}{2\lambda}
     \nonumber\\
    &&+ \frac{((E^x)')^2}{(E^\varphi)^2} \left(- 
       \frac{E^{\varphi}}{4E^x}   
    - i \frac{\lambda}{E^x}\left(E^xP_x-\frac{1}{2}
       E^{\varphi}P_{\varphi}   \right)  \exp(2i
       \lambda P_\varphi)
    \right)
    \nonumber\\
    &&
    + \left( \frac{(E^x)'
       (E^\varphi)'}{(E^\varphi)^2}
    - \frac{(E^x)''}{E^\varphi} \right) \exp ( 2i\lambda
       P_\varphi ) \Bigg]
\end{eqnarray}
with $\lambda\propto 1/\sqrt{E^x}$ and implies a classical structure function.
In our notation, this model can be realized as follows: We first perform the
inverse of the canonical transformation (\ref{canonicalx}), replacing the
function $\lambda$ with a constant $\bar{\lambda}$ and identifying
$E^xP_x-\frac{1}{2}E^{\varphi}P_{\varphi}$ as the new momentum of $E^x$,
resulting in the transformed constraint
\begin{eqnarray}\label{HComplex-periodic}
    H[N]
    &=& -\frac{1}{2G}\int{\rm d}x N \frac{\bar{\lambda}}{\lambda} 
 \sqrt{E^x} \Bigg[
   \frac{E^\varphi}{E^x} \left(\frac{\lambda^2}{\bar{\lambda}^2} + 3 \frac{\sin^2 (\bar{\lambda} P_\varphi)}{\bar{\lambda}^2}\right) 
    + 4 P_x
   \frac{\sin (2 \bar{\lambda} P_\varphi)}{2\bar{\lambda}}
     \\
    &&+ \frac{((E^x)')^2}{(E^\varphi)^2} \left(- \frac{3E^\varphi}{4 E^x}   
    - i \bar{\lambda} P_x 
    \right) \exp(2i\bar{\lambda} P_\varphi)
    + \left( \frac{(E^x)'(E^\varphi)'}{(E^\varphi)^2}
    - \frac{(E^x)''}{E^\varphi} \right) \exp ( 2i\bar{\lambda} P_\varphi) \Bigg]\,.\nonumber
\end{eqnarray}
The classical structure function of the original model is then transformed to 
\begin{equation}\label{StructureComplex-periodic}
    \tilde{q}^{xx}=\frac{\bar{\lambda}^2}{\lambda^2} \frac{E^x}{(E^\varphi)^2}\,.
\end{equation}

The new Hamiltonian constraint is strictly periodic in $P_{\varphi}$.  Since
the specific coefficients in 
(\ref{HComplex}) and (\ref{HComplex-periodic}) turn out to require
non-trivial values for all the modification functions in (\ref{H}), we will
first demonstrate how it is possible to obtain a classical-type structure
function (\ref{StructureComplex-periodic}), which is independent of $(E^x)'$
and $P_{\varphi}$, even in the case of modified dynamics.

To this end, we apply a second canonical transformation that replaces
$P_{\varphi}$ with $P_{\varphi}+\mu_{\varphi}$ where $\mu_{\varphi}$ is
constant, leaving all other phase-space variables unchanged. (More generally,
one may use this $K_{\varphi}$-transformation with an $E^x$-dependent function
$\mu_{\varphi}$, in which case the other phase-space variables change. Such a
transformation has been used in one of the intermediate steps of
\cite{HigherCov}.)  We choose
$\chi=\sqrt{2}\tilde{\chi} (\bar{\lambda}/\lambda)
\exp(-i\bar{\lambda}\mu_{\varphi})$ and
$c_f=\tilde{c}_f\exp(2i\bar{\lambda}\mu_{\varphi})$ in (\ref{H}). This choice
renders the constraint complex. In the final step, we will replace
$\mu_{\varphi}$ with $-i\epsilon$ with a real $\epsilon$ and take the limit
$\epsilon\to\infty$.

In the general emergent metric (\ref{emergent}), the diverging limiting
contribution to $c_f$ dominates the term
$\bar{\lambda}^2((E^x)')^2/(E^{\varphi})^2$. The limiting emergent metric is
therefore independent of $(E^x)'$. If we combine the $c_f$-part of the
emergent metric with a non-zero $q$-term, choosing
$q=\frac{1}{2}ic_f/\bar{\lambda}$, we obtain
\begin{eqnarray}
 && \left(\left(c_{f}
    + \left(\frac{\bar{\lambda} (E^x)'}{2 E^\varphi} \right)^2 \right) \cos^2 \left(\bar{\lambda} (P_\varphi+\mu_{\varphi})\right)
    - 2 q \bar{\lambda}^2 \frac{\sin \left(2 \bar{\lambda} (P_\varphi+\mu_{\varphi})\right)}{2
    \bar{\lambda}}\right)\chi^2\nonumber\\
  &\longrightarrow & \tilde{c}_f
      \left(1+\cos(2\bar{\lambda}(P_{\varphi}+\mu_{\varphi}))
      -i \sin (2
         \bar{\lambda} (P_\varphi+\mu_{\varphi}))\right)
      \tilde{\chi}^2\frac{\bar{\lambda}^2}{\lambda^2} \longrightarrow
                     \tilde{c}_f\tilde{\chi}^2
                     \frac{\bar{\lambda}^2}{\lambda^2} 
\end{eqnarray}
where arrows indicate limiting values for $\mu_{\varphi}=-i\epsilon$ with
$\epsilon\to\infty$. The $P_{\varphi}$-dependence then also disappears.
Choosing $\tilde{c}_f=1/\tilde{\chi}^2$, we obtain an emergent metric that is
identical with the classical metric up to a factor of
$\bar{\lambda}^2/\lambda^2$. This factor can be removed by using suitable
$E^x$-dependent $\tilde{c}_f$ or $\tilde{\chi}$, or by applying the inverse of
(\ref{canonicalx}) and transforming back to a constraint that is not periodic
in $P_{\varphi}$. Even though the emergent metric is then identical with the
classical metric, the uniqueness results of \cite{Regained} are circumvented
in this case by working with dynamics that is not time-reversal symmetric.

\subsubsection{Comparison}

In \cite{MassCovariance}, the classical metric had also been obtained, but
with several other coefficients in (\ref{HComplex}) that imply
non-trivial choices of the remaining modification functions. In order to embed
this model in emergent modified gravity, we first
redefine some of the general modification functions according to
\begin{eqnarray}
  \chi&=&\tilde{\chi} \exp(-2i\bar{\lambda}\mu_\varphi)\\
 c_f&=&\frac{1}{2}i \bar{\lambda} (c_2 -c_1\exp(4i\bar{\lambda}\mu_\varphi))\\
q&=&\frac{1}{4}i (c_2 +c_1\exp(4i\bar{\lambda}\mu_\varphi))\\
V&=&\frac{1}{4}\tilde{V}
     \exp(2i\bar{\lambda}\mu_{\varphi})+\frac{2\sqrt{E^x}}{\bar{\lambda}}
     \frac{\exp(4 
    i \bar{\lambda} \mu_\varphi)}{2i} \left(\frac{\alpha c_1}{2E^x} + \frac{\partial
     c_1}{\partial E^x} \right)
\end{eqnarray}
such that
\begin{eqnarray}\label{H-complex}
    H[N]
    &\longrightarrow & -\frac{1}{2G}\int{\rm d}x N \frac{\tilde{\chi}}{4}
                       \sqrt{E^x} \Bigg[ E^\varphi \Bigg( 
    -\frac{\tilde{V}}{2\sqrt{E^x}}
    \nonumber\\
    &&
    + \frac{4}{\bar{\lambda}} \left(\frac{e^{2i\bar{\lambda} P_\varphi}}{2i} \left(\frac{\alpha c_2}{2E^x} + \frac{\partial c_2}{\partial E^x} \right)-\frac{e^{-2i\bar{\lambda} P_\varphi}}{2i} \left(\frac{\alpha c_1}{2E^x} + \frac{\partial c_1}{\partial E^x} \right)\right)
    \Bigg)
    \nonumber\\
    &&
    + 4 P_x \left(c_2 e^{2i\bar{\lambda}P_\varphi}+c_1 e^{-2i\bar{\lambda}P_\varphi}\right)
    \nonumber\\
    &&
    + \frac{((E^x)')^2}{(E^\varphi)^2} \left(
    - \frac{\alpha E^\varphi}{4 E^x}
    - i \bar{\lambda} P_x
    \right) e^{2i\bar{\lambda} P_\varphi}
    \nonumber\\
    &&
    + \left( \frac{(E^x)' (E^\varphi)'}{(E^\varphi)^2}
    - \frac{(E^x)''}{E^\varphi} \right) e^{2i\bar{\lambda} P_\varphi}
       \Bigg]\,,
\end{eqnarray}
while the structure function (\ref{emergent}) becomes
\begin{eqnarray}
    \tilde{q}^{xx} &=& \frac{i\bar{\lambda}}{2} \Bigg(
    c_2 e^{i \bar{\lambda} P_\varphi-3\bar{\lambda}\epsilon} \cos \left(\bar{\lambda} (P_\varphi-i\epsilon)\right)
    - \frac{c_1}{2} \left(1+e^{-2i\bar{\lambda} P_\varphi-2\bar{\lambda}\epsilon}\right)\Bigg)
    \tilde{\chi}^2 \frac{E^x}{(E^\varphi)^2}
    \nonumber\\
    &\longrightarrow & - \frac{i\bar{\lambda}\tilde{\chi}^2 c_1}{4}
    \frac{E^x}{(E^\varphi)^2}\,.
\end{eqnarray}

The constraint (\ref{HComplex-periodic}) and the structure function
(\ref{StructureComplex-periodic}) are recovered in the special case where
$\tilde{\chi}=4\bar{\lambda}/\lambda$, $c_2=1/(4i\bar{\lambda})$,
$c_1=-1/(4i\bar{\lambda})$, $\alpha=3$, and
$\tilde{V}\sqrt{E^x} = - (3+2 \lambda^2)/\bar{\lambda}^2$. In particular, for
$\lambda\propto 1/\sqrt{E^x}$, the dilaton potential contains not only a
contribution of the form realized in vacuum general relativity, where
$V\propto 1/\sqrt{E^x}$, but also a term proportional to $(E^x)^{-3/2}$. This
model is therefore a rather contrived combination of different types of
modifications, including choices of the dilaton potential and (complex)
holonomy-like terms, as well as other implications implied by non-classical
choices of $\chi$, $c_f$, $q$ and $\alpha$. A reliable analysis of
corresponding physical effects is possible only if these different types of
modifications are unraveled by embedding the model of \cite{MassCovariance}
within emergent modified gravity. Without this connection, the model has the
status of a mathematical curiosity for which a simple solution of the equation
relating the mass observable to a valid structure function happens to be
possible without offering physical motivations for this particular choice. We
will present further comparisons between these methods and emergent modified
gravity in Section.~\ref{s:Conclusions}.

One may question the appearance of complex terms in the Hamiltonian
constraint, even if the emergent metric is real. General equations of motion
and non-static solutions could then also be complex. However, it is easy to
see that redefining $P_x\to iP_x$ and $P_{\varphi}\to i P_{\varphi}$ renders
all terms real, while it replaces trigonometric functions of the old
$P_{\varphi}$ with hyperbolic functions of the new
$P_{\varphi}$. (Alternatively, all terms are real if $\bar{\lambda}$ is purely
imaginary, such that it can be replaced with a real $\mu=i\bar{\lambda}$.) The
possible appearance of hyperbolic functions in emergent modified gravity has
been pointed out in \cite{HigherCov}, where it is related to a sign parameter
$s$ in the general classification. Since multiplying only the momenta by $i$
is not a canonical transformation, this model is physically distinct from
(\ref{Hlambda}) even though it may appear to be based on holonomy
modifications if one does not pay attention to the complex nature of some
terms. A physical implication in a related version, in which the emergent
metric is not classical but hyperbolic functions of $P_{\varphi}$ have also
been used, is large-scale signature change as analyzed in \cite{EmergentSig}.

\section{Conclusions}
\label{s:Conclusions}

Emergent modified gravity determines consistent and generally covariant
gravitational theories within a given range of effective field theory,
delineated for instance by symmetry conditions or spatial derivative
orders. It goes beyond traditional classifications of modified gravity from an
action viewpoint because it eliminates the assumption that equations of motion
are formulated directly for a space-time metric field or alternatives such as
a tetrad or a connection in first-order formulations. Instead, consistency and
covariance conditions are formulated canonically, such that a compatible
space-time metric emerges from properties of the field equations.

The canonical setting and the crucial role of covariance conditions imply that
emergent modified gravity, for specific choices of modification functions, is
an important ingredient in consistent formulations of models of loop quantum
gravity. In this context, the classical canonical equations are modified in
ways motivated by certain operators used on the kinematical Hilbert space
proposed by loop quantum gravity. The resulting class of modification
functions are of a form in which the classical definition of the space-time
metric in terms of gravitational phase-space variables is no longer compatible
with the modified Hamiltonian constraint: Gauge transformations of the
modified theory do not agree with coordinate transformations of the classical
metric, and therefore the classical metric does not constitute a
gauge-invariant characterization of geometrical properties of solutions. In
some cases, if specific relationships between different modification functions
are fulfilled, an emergent space-time metric exists that does provide
gauge-invariant geometrical information for solutions, making the theory
generally covariant. This emergent metric is derived within emergent modified
gravity, which also determines restrictions on modification functions such
that the covariance condition is fulfilled.

It is therefore essential that the emergent space-time metric be taken into
account for reliable geometrical interpretations, which has not been done in
most models of loop quantum gravity such as the recent
\cite{MimeticCGHS,LTBShock,LTBPol,LTBPolII}. Generically, the spatial metric
receives additional corrections compared with the lapse function, such that
the classical relationship $N^2=q^{xx}$ in a Schwarzschild geometry no longer
holds. Also this implication has been overlooked in models of loop quantum
gravity, which often try to construct space-time line elements that might
modify the radial dependence of $N^2$ and $q^{xx}$ but not their classical
equality, as for instance in \cite{HolOppSny}.

Covariant models of gravitational collapse that are compatible with emergent
modified gravity have been formulated and analyzed in \cite{EmergentFluid} as
well as \cite{SphSymmLTB1,SphSymmLTB2}, using earlier constructions from
\cite{SphSymmEff,SphSymmEff2} that amount to a constant holonomy function as
seen from our present discussion. The recent \cite{MassCovariance} constructs
two examples of spherically symmetric models with scale-dependent $\lambda$
that turn out to be special cases of the general classification given in
\cite{HigherCov,EmergentScalar}, corresponding to the models given here in
Sections~\ref{s:Universal} (with $\lambda\propto 1/\sqrt{E^x}$) and
\ref{s:TimeReversal}, respectively. In this approach, the starting point is a
modified mass observable which is related to the Hamiltonian constraint and to
the emergent metric through two independent partial differential
equations. Making a suitable choice for the mass observable then allows one to
derive a compatible metric, as well as the corresponding Hamiltonian
constraint. However, this paper did not perform a complete covariance
analysis. Instead, a sufficient but not necessary condition for covariance was
stated in the property that the resulting emergent metric depend only on $E^x$
and the mass observable. This requirement imposes an implicit condition on
allowed choices of mass functions that has not been systematically
evaluated. Accordingly, the methods of \cite{MassCovariance} can only lead to
examples of modified covariant theories derived from consistent choices of
mass functions, but they are not suitable for a general classification of
covariant theories as done in emergent modified gravity. Moreover, these
methods do not pay attention to the derivative order of different terms in the
Hamiltonian constraint. From the perspective of gravity as an effective field theory, it is
therefore unclear whether new effects are reliable or could be altered or
eliminated if other terms of the same derivative order are included (from
emergent modified gravity or higher curvature terms) in the
constraint. Emergent modified gravity has classified all covariant theories
(so far) up to second order in spatial derivatives, and it has independently
derived compatible mass observables that are indeed restricted in their
functional dependence on the phase-space variables \cite{EmergentScalar}. No
such restriction has been given in \cite{MassCovariance}. Finally, relying on
the condition that the emergent metric depend only on $E^x$ and the mass
observable makes the approach unsuitable for a general inclusion of matter
terms, where the mass dependence is no longer realized.

Compared with these formulations, emergent modified
gravity is covariant not only in space-time but also in phase space because it
derives equivalence classes of compatible modified theories and their emergent
metrics under canonical transformations of the gravitational or matter
variables. This additional ingredient is important because the modified
dynamics implies that phase-space functions such as $E^x$ and
$(E^{\varphi})^2/E^x$ do not retain their classical relationship with
(emergent) metric components, and the momenta $P_x$ and $P_{\varphi}$ have
non-classical relationships with components of extrinsic curvature. In the
absence of a-priori geometrical interpretations of phase-space degrees of
freedom, which is an essential aspect of emergent space-time theories, there
is no preferred choice of canonical variables. Entire equivalence classes of
modified theories under canonical transformations should therefore be
considered. This phase-space covariance allowed us to present a full
discussion of different holonomy schemes in models of loop quantum gravity,
going beyond the examples provided in \cite{SphSymmLTB1,SphSymmLTB2,MassCovariance}.

In the present paper, we have presented a discussion in this spirit for
holonomy modifications in models of loop quantum gravity, which are often
considered the most important non-classical effect in this theory. We have
shown by several examples as well as general expressions that these terms in a
Hamiltonian constraint can be completely understood only if properties of
emergence and covariance, in both the space-time and phase-space meanings, are
taken into account. We have arrived at a unified treatment of previous
approaches that had traditionally been separated as distinct modification
schemes, but are realized now as different choices of canonical variables and
modification functions. A complete description of holonomy terms
requires a multitude of modification functions that transform non-trivially
under canonical transformations of the basic phase-space variables. This set
of functions replaces the single holonomy function referred to in traditional
schemes.

We have therefore reconciled previous results that seemed to rule out certain
choices of holonomy modifications, in particular those of $\mu_0$-type with a
constant holonomy length. Such expressions are easier to quantize in terms of
basic holonomy operators, at least in spherically symmetric models, but taken
in isolation they often seem to imply large modifications in semiclassical
regimes such as late-time cosmology. We have demonstrated that the inclusion
of the full set of modification functions in a covariant setting, as well as
canonical transformations between different holonomy schemes, makes models
with constant holonomy functions consistent with classical late-time or
asymptotic behavior. This result suggests that loop quantized versions of
these models should be less involved than it appears in constructions based on
non-constant holonomy functions.

Moreover, a considerable number of models that have been proposed is
incompatible with the existence of a geometrical interpretation of their
solutions because they violate covariance conditions. Models of quantum
gravity are expected to imply some kind of quantum space-time, such as
discrete versions, in which classical covariance need not be realized
strictly. However, physical evaluations of models of loop quantum gravity in
cosmological or black-hole applications make use of effective line elements
that implicitly assume a compatible (pseudo-)Riemannian geometry. Such line
elements are then used for further derivations of standard space-time
properties, such as curvature tensors, geodesics, or horizons. At this point,
covariance conditions become important because reliable geometrical properties
should be independent of gauge or deparameterization choices made in the
formulation of a modified theory. This is precisely the main condition
analyzed by emergent modified gravity. If the resulting covariance conditions
are violated, the assumed effective description by a line element is not gauge
invariant and therefore does not describe physical implications of the
underlying theory. Importantly, emergent modified gravity has shown that
anomaly-freedom of the gravitational constraint is a necessary but not
sufficient condition for covariance in this broad sense. Examples of
anomaly-free but non-covariant models are those constructed in
\cite{LTBPol,LTBPolII,LTBShock}, while those constructed in \cite{MimeticCGHS}
are not even anomaly-free. (See also \cite{Mimetic}.)

Another important implication of emergent modified gravity is therefore that it
leads to strong conditions that can be used to eliminate ambiguities and choices
in current formulations of models of loop quantum gravity, while maintaining
enough freedom for different holonomy schemes to be realized. Combined with the
emergent space-time metrics implied by this theory, reliable space-time
implications can be derived in models of loop quantum gravity.

\section*{Acknowledgements}

IHB is supported by the Indonesia Endowment Fund for Education (LPDP) grant
from the Ministry of Indonesia. SB is supported in part by the Higgs Fellowship and by the STFC Consolidated
Grant ``Particle Physics at the Higgs Centre.'' The work of MB and ED was
supported in part by NSF grant PHY-2206591.


\end{document}